\documentclass[12pt,a4paper]{article}

\usepackage{amsmath}
\usepackage{amssymb}
\usepackage{graphicx}
\usepackage{dsfont}
\usepackage{overpic}
\usepackage{cite}

\let\oldappendix=\appendix
\let\oldsection=\section
\renewcommand{\appendix}{\oldappendix%
\def\theequation{\Alph{section}.\arabic{equation}}%
\renewcommand{\section}{\setcounter{equation}{0}\oldsection}}

\newcommand{\vs}{\vspace{-0.20cm}}
\newcommand{\beq}{\begin{equation}}
\newcommand{\eeq}{\end{equation}}
\newcommand{\beqa}{\begin{eqnarray}}
\newcommand{\eeqa}{\end{eqnarray}}
\newcommand{\no}{\nonumber}

\newcommand{\tr}{\mbox{tr}}

\newcommand{\newop}[2]{\def#1{\mathop{\mathrm{#2}}\nolimits}}
\newop{\artanh}{artanh}
\newop{\det}{det}
\newop{\tr}{tr}
\newop{\diag}{diag}
\newop{\Re}{Re}
\newop{\Im}{Im}

\newcommand{\Lagr}{\mathcal{L}}

\newcommand{\slashed}[1]{{#1} \! \! \! /}

\begin{document}

\hfill 

\hfill 

\bigskip\bigskip

\begin{center}

{{\Large\bf  Chiral corrections to the Roper 
mass\footnote{This
research is part of the EU Integrated Infrastructure Initiative Hadron Physics Project
under contract number RII3-CT-2004-506078. Work supported in part by DFG (SFB/TR 16,
``Subnuclear Structure of Matter'', and BO 1481/6-1) and by Canada's NSERC and CRC programs.}}}

\end{center}

\vspace{.4in}

\begin{center}
{\large B.~Borasoy\footnote{email: borasoy@itkp.uni-bonn.de}$^a$,
        P.~C.~Bruns\footnote{email: bruns@itkp.uni-bonn.de}$^a$,
        U.-G.~Mei{\ss}ner\footnote{email: meissner@itkp.uni-bonn.de}$^{a,b}$,
        R.~Lewis\footnote{email: randy.lewis@uregina.ca}$^c$}

\bigskip

\bigskip

$^a$Helmholtz-Institut f\"ur Strahlen- und Kernphysik (Theorie) \\
Universit\"at Bonn, Nu{\ss}allee 14-16, D-53115 Bonn, Germany \\[0.3cm]
$^b$Institut f\"ur Kernphysik (Theorie), Forschungszentrum J\"ulich \\
D-52425 J\"ulich, Germany \\[0.3cm]
$^c$Department of Physics, University of Regina, Regina, Saskatchewan, Canada,
S4S 0A2\\

\vspace{.2in}

\end{center}

\vspace{.7in}

\thispagestyle{empty} 

\begin{abstract}\noindent
We analyze the quark mass dependence of the Roper mass to one-loop order 
in relativistic baryon chiral 
perturbation theory. The loop integrals are evaluated using infrared regularization
which preserves chiral symmetry and establishes a chiral counting scheme.
The derived chiral expansion of the Roper mass may prove useful for chiral
extrapolations of lattice data. For couplings of natural size 
the quark mass dependence
of the Roper mass is similar to the one of the nucleon.
\end{abstract}\bigskip

\begin{center}
\begin{tabular}{ll}
\textbf{PACS:}& 12.38.Gc, 12.39.Fe, 12.40.Yx \\[6pt]
\textbf{Keywords:}& Lattice QCD calculations, chiral Lagrangians, 
hadron mass models and calculations
\end{tabular}
\end{center}


\vfill\eject

\section{Introduction}\label{sec:Intro}
Understanding the (ir)regularities of the light quark baryon spectrum poses an 
important challenge for lattice QCD. In particular, the first even-parity
excited state of the nucleon, the Roper $N^* (1440)$ (from here on called the
Roper) is very intriguing---it is lighter than the first odd-parity nucleon
excitation, the $S_{11} (1535)$, and also has a significant branching ratio 
into two pions. Recent lattice studies, see e.g.
\cite{Sasaki:2001nf,Guadagnoli:2004wm,Leinweber:2004it,Mathur:2003zf,Burch:2006cc,
Sasaki:2005ap,Sasaki:2005ug},
have not offered a clear picture about the nucleon resonance spectrum. 
In particular, lattice 
QCD operates at unphysical quark (pion) masses and thus a chiral
extrapolation is needed to connect these data to the physical world. The
findings of  Ref.~\cite{Mathur:2003zf} indicate a rapid cross over of the
first positive and negative excited nucleon states close to the chiral
limit. No such level switching is e.g. found in ~\cite{Guadagnoli:2004wm},
possibly related to the fact that the simulations were performed at quark
masses too far away from the chiral regime. It should also be noted that so far
very simple chiral extrapolation 
functions have been employed in most approaches, e.g., a linear 
extrapolation in the quark masses, 
thus  $\sim m_\pi^2$ (with $m_\pi$ the pion mass), was
applied in \cite{Burch:2006cc}. It is therefore important to provide the
lattice practitioners with improved chiral extrapolation functions. This
is the aim of this manuscript. We consider the Roper mass (the real part
of the Roper self--energy) to one-loop in baryon chiral perturbation theory.
More precisely, we employ an extension of the infrared regularization method 
of Ref.~\cite{Becher:1999he} and study the pion mass dependence as a function
of the various low-energy (coupling) constants that appear in the expression.
We refrain from analyzing the existing lattice data---our results apply to
full QCD and not to one of the various approximations to QCD employed in the
lattice studies.

\medskip\noindent
Our manuscript is organized as follows. In Sec.~\ref{sec:Leff}, we display 
the effective chiral Lagrangian underlying our calculation. The chiral 
corrections to the Roper mass are calculated in Sec.~\ref{sec:Rmass}. 
Sec.~\ref{sec:qmasdep} contains our  results and the discussion thereof.

\section{Effective Lagrangian}\label{sec:Leff}
We will calculate chiral corrections to the Roper mass up to one-loop order. 
Since the Roper is the first even-parity excited state of the nucleon, the
construction of the chiral SU(2) effective Lagrangian follows standard
procedures,  see e.g.~\cite{Fettes:2000gb}. The effective Lagrangian 
relevant for our  calculation is (see also Ref.~\cite{Borasoy:1996bx})
\beq  \label{eq:Lagrangian}
\Lagr = \Lagr_0 + \Lagr_{R} + \Lagr_{NR}~,
\eeq
with the free part
\beq
\Lagr_0 = i \bar{N} \gamma_\mu D^\mu N - M_N \bar{N} N 
     +  i \bar{R} \gamma_\mu D^\mu R - M_R \bar{R} R ~, 
\eeq
where $N, R$ are nucleon and Roper fields, respectively, and $M_N, M_R$
the corresponding baryon masses in the chiral limit. 
$D_\mu$ is the chiral covariant derivative, for our purpose we can set
$D_\mu = \partial_\mu$, see e.g.~\cite{Fettes:2000gb} for definitions. 
The pion-Roper coupling  is given  to leading chiral order by
\beq
\Lagr_{R}^{(1)} =  \frac{1}{2} g_R \bar{R} \gamma_\mu \gamma_5 u^\mu R
\eeq
with an unknown coupling $g_R$ and the superscript denotes the chiral order. 
The pion fields are collected in $u_\mu = - \partial_\mu 
\mbox{\boldmath$ \pi$} / f_\pi 
+ {\cal O}(\mbox{\boldmath$ \pi$}^3)$,
where $f_\pi$ is the pion decay constant in the chiral limit.
At next--to--leading order, the relevant terms in $\Lagr_{R}$ are 
(we work in the isospin limit $m_u = m_d$ and neglect electromagnetism)
\beq
\Lagr_{R}^{(2)} =  c_1^* \langle \chi_+ \rangle \bar{R}  R  
               - \frac{c_2^*}{8 M_R^2}  \bar{R} \left(\langle u_\mu u_\nu \rangle 
                    \{D^\mu, D^\nu \} + {\rm h.c.} \right)  R
               + \frac{c_3^*}{2}  \langle u_\mu u^\mu \rangle \bar{R}  R~ \ ,
\eeq
where $\chi_+ $ is proportional to the pion mass and induces explicit chiral 
symmetry breaking.  Further, $\langle ~ \rangle$ denotes the trace in 
flavor space.
For a complete one loop calculation we also need the fourth  order effective
Lagrangian, more precisely, the term
\beq
\Lagr_{R}^{(4)} =  - \frac{e_1^*}{16} \langle \chi_+ \rangle^2 \bar{R}  R  \ .
\eeq
The interaction piece between nucleons and the Roper reads
\beq
 \Lagr_{NR}^{(1)} = \frac{1}{2} g_{NR} \bar{R} \gamma_\mu \gamma_5 u^\mu N 
+ {\rm h.c.} \ .
\eeq
The coupling $g_{NR}$ can be determined from the strong decays of the
resonance $R$, its actual value is given below.
In principle a term of the form
\beq
i  \lambda_1 \bar{R} \gamma_\mu D^\mu N - \lambda_2 \bar{R} N + {\rm h.c.} 
\eeq
is possible, but applying the equations of motion removes the first term
 (and its hermitian conjugate)
such that we are left with the terms $\bar{R} N$ and $\bar{N} R$. These terms induce
mixing between the nucleon and Roper fields, but diagonalization of the $N$-$R$
mass matrix does not lead to new operator structures and its effect can be completely
absorbed into the couplings already present in the Lagrangian. 
We can thus safely work with the Lagrangian in Eq.~(\ref{eq:Lagrangian}).
A complete one-loop calculation involves tree graphs with insertion of chiral
dimension two and four and one-loop graphs with at most one insertion from
${\cal L}_{R}^{(2)}$.

\section{Chiral corrections to the Roper mass}\label{sec:Rmass}

We are now in the position to work out the various contributions to the
Roper mass. The loop diagrams are evaluated making use of (an extension of)
the infrared regularization (IR) method \cite{Becher:1999he}.
As we will see, the IR scheme is suited for the study of systems with one light
mass scale $m_\pi$ and two heavy mass scales $M_N, M_R$ with $M_N^2 \ll
M_R^2$. In the real world, we have $M_R^2/M_N^2 \simeq 2.4$, so that this
condition is approximately fulfilled.

\smallskip\noindent
{\bf 1. Tree level:}
Only the $c_1^*$ and  $e_1^*$ terms contribute to the self-energy at tree level
\beq
\Sigma_R^{\rm tree} = - 4 c_1^* \, m_\pi^2 + e_1^* \, m_\pi^4 ~.
\eeq
These terms could be absorbed into $M_R$, but since we are interested in the explicit
dependence on the pion (quark) mass, we must keep them. The first
term is the leading order contribution to the nucleon-Roper $\sigma$-term.

\begin{figure}[t]
\centering
\includegraphics[width=0.7\textwidth]{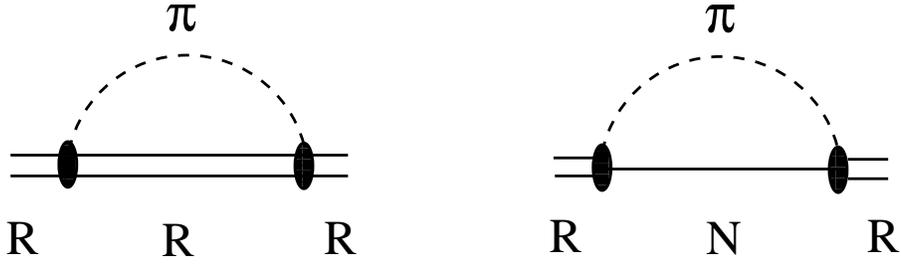} 
\caption{One-loop self-energy graphs of the Roper (R) with intermediate
Roper-pion ($\pi$)  and nucleon (N)-pion states, respectively.
}
\label{fig:roperself}
\end{figure}

\smallskip\noindent
{\bf 2. Pion-nucleon loop:}
This is the only new structure compared e.g. to the calculation of
the nucleon self-energy, see Fig.~\ref{fig:roperself}. We extend here the 
method of Ref.~\cite{Bruns:2004tj} developed for IR with vector mesons. 
Consider first the fundamental scalar integral at one-loop order in $d$ dimensions
with an intermediate pion-nucleon pair and external momentum $p$
\beq  \label{eq:scalarloop}
I_{\pi N} (p^2)  =  \int \frac{d^d l}{(2 \pi)^d}  \ \frac{i}{[l^2 - m_\pi^2 + i \epsilon] 
\ [(p+l)^2 - M_N^2 + i \epsilon]} \ .
\eeq
Employing standard Feynman parameterization leads to 
\beq  \label{eq:parint}
I_{\pi N} (p^2)  = - \frac{\Gamma (2 - \frac{d}{2})}{(4 \pi)^{d/2}} M_N^{d-4}
  \int_0^1 dz \ \left( \beta [z - x_+] [z - x_-]   \right)^{\frac{d}{2}-2}
\eeq
in terms of the parameters
\beqa
x_{\pm} &=&  \frac{\alpha + \beta -1}{2 \beta} \left( 1 \pm 
         \sqrt{1 - \frac{4 \alpha \beta}{(\alpha + \beta -1)^2} }  \  \right)~, \quad
\alpha = \frac{m_\pi^2}{M_N^2} ~, \quad
\beta = \frac{p^2}{M_N^2}~.
\eeqa
In the following, we will restrict ourselves to the kinematical region
$p^2 \gg (M_N +m_\pi)^2$ which is equivalent to
\beq
\frac{4 \alpha \beta}{(\alpha + \beta -1)^2} \ll 1~.
\eeq
This implies that in the chiral limit $\alpha \to 0$ the mass difference $p^2 - M_N^2$
remains finite and does not tend towards zero as in the standard case of IR.
This constraint is clearly satisfied for values $p^2 \approx M_R^2$ 
close to the Roper mass. It is also consistent with resonance decoupling in
the chiral limit \cite{Gasser:1979hf}.
The nucleon propagator is thus counted as zeroth chiral order for momenta $p^2 \approx M_R^2$.
Setting $\beta = M_R^2/M_N^2$ one obtains the small parameter
\beq
\frac{4 \alpha \beta}{(\alpha + \beta -1)^2} \approx \frac{1}{9}~,
\eeq
which indicates a fast convergence of the expansion of the loop 
integral~Eq.~(\ref{eq:scalarloop}) 
in powers of the pion mass. Expansion of $x_{\pm}$ in $\alpha$ leads to
\beqa
x_+ &=& \frac{\beta -1}{\beta} - \frac{\alpha}{\beta(\beta -1)} 
- \frac{\alpha^2}{(\beta -1)^3} 
         + {\cal O}(\alpha^3) \ , \no \\
x_- &=& \frac{\alpha}{\beta -1} + \frac{\alpha^2}{(\beta -1)^3} 
         + {\cal O}(\alpha^3) \ ,
\eeqa
where $x_- = {\cal O} (\alpha)$ and $x_+ = {\cal O} (1)$.
We now divide the parameter integral of Eq.~(\ref{eq:parint}) into three parts
\beq
I_{\pi N} = - \frac{\Gamma (2 - \frac{d}{2})}{(4 \pi)^{d/2}} M_N^{d-4}
           \left( I_{\pi N}^{(1)} + I_{\pi N}^{(2)} + I_{\pi N}^{(3)}  \right)~,
\eeq
with 
\beqa  
I_{\pi N}^{(1)} (p^2)  &=& \int_0^{x_-} dz \ \left( \beta [z - x_+] [z - x_-]   
                                   \right)^{\frac{d}{2}-2}~, \no \\
I_{\pi N}^{(2)} (p^2)  &=& \int_{x_-}^{x_+} dz \ \left( \beta [z - x_+] [z - x_-] 
                                       \right)^{\frac{d}{2}-2}~, \no \\
I_{\pi N}^{(3)} (p^2)  &=& \int_{x_+}^1 dz \ \left( \beta [z - x_+] [z - x_-]   
\right)^{\frac{d}{2}-2} ~.
\eeqa
Note  that $0 < x_- < x_+ < 1$.
The first integral $I_{\pi N}^{(1)}$ can be rewritten as
\beq
I_{\pi N}^{(1)} (p^2) = (x_-)^{d/2-1} \int_0^1 dy \ \left( \beta [x_+ - x_- (1-y)]   
                           \right)^{\frac{d}{2}-2} y^{\frac{d}{2}-2} \ .
\eeq
Expansion of the integrand in powers of $x_- \sim {\cal O}(\alpha)$ 
and interchanging summation with integration leads to
\beq
I_{\pi N}^{(1)} (p^2) = \frac{2}{d-2} (x_-)^{d/2-1} ( \beta x_+ )^{d/2-2}
         + {\cal O}(\alpha^{d/2}) \ ,
\eeq
where the leading term is of order ${\cal O}(\alpha^{d/2-1})$ and thus conserves
power counting. The integral
$I_{\pi N}^{(1)}$ contains only fractional powers of $\alpha$ and  contributes 
to the infrared singular part. The remaining two integrals,
on the other hand, are regular in $\alpha$. For $I_{\pi N}^{(2)}$ one has
\beq
I_{\pi N}^{(2)} (p^2) = \left( - \beta \right)^{d/2-2}
           \int_{x_-}^{x_+} dz \ \left(  [x_+-z] [z - x_-] \right)^{\frac{d}{2}-2} \ ,
\eeq
where $\beta$ has a small positive imaginary piece. 
Integration yields
\beq
I_{\pi N}^{(2)} (p^2) = \left( - \beta \right)^{d/2-2} (x_+ -x_-)^{d-3} \frac{\left(\Gamma 
              (\frac{d}{2}-1)\right)^2}{\Gamma (d-2)} \ .
\eeq
Obviously, this expression is expandable in powers of $x_-$. 
The integral $I_{\pi N}^{(2)}$ is complex
and does not conserve power counting.
It contributes only to the
regular part and will be omitted for our purposes. 
More precisely, the imaginary part does not contribute to the resonance mass,
while the real part can be absorbed into the couplings of the effective Lagrangian
at the on-shell momentum $p^2 =M_R^2$.
In the third integral,
\beq  
I_{\pi N}^{(3)} (p^2)  = \int_{x_+}^1 dz \ \left( \beta [z - x_+] [z - x_-] 
  \right)^{\frac{d}{2}-2} \ ,
\eeq
one can expand the integrand directly in powers of $x_-$.
The expansion coefficients of this series are integrals of the type $(r \in \mathds{R})$
\beq
\int_{x_+}^1 dz \ \left( z - x_+  \right)^{\frac{d}{2}-2} z^r
= \left( 1 -x_+ \right)^{d/2-1} \int_{0}^{1} dw 
      \ \left( 1 - w  \right)^{\frac{d}{2}-2} 
              \left(1 + w (x_+ -1) \right)^r \ .
\eeq
Since $1-x_+$ remains finite in the chiral limit these integrals also 
contribute only to the regular part and can be absorbed into the
couplings of the Lagrangian at momentum $p^2 =M_R^2$. 
Therefore, the infrared singular part which stems from small values of the Feynman 
parameter $z$ is entirely contained in $I_{\pi N}^{(1)}$ and 
we can restrict ourselves to the integral
\beqa
(I_{\pi N})_{IR} &=& - \frac{\Gamma (2 - \frac{d}{2})}{(4 \pi)^{d/2}} M_N^{d-4}
     \int_0^{x_-} dz \ \left( \beta [x_+-z] [x_--z]  \right)^{\frac{d}{2}-2} \no \\
&=&
    - \frac{M_N^{d-4}}{16 \pi^2} \left\{ x_- \left( \frac{2}{4-d} 
    +  \ln 4\pi - \gamma_E +1  \right) -
     \int_0^{x_-} dz \ \ln \left( \beta [x_+-z] [x_--z]  \right) \right\} \, . 
\eeqa
Expansion in $\alpha$ leads to
\beq  \label{eq:IRresult}
(I_{\pi N})_{IR} = \left( 2 L + \frac{1}{16 \pi^2} \ln 
\left( \frac{m_\pi^2}{M_R^2}\right) 
\right) \left( \frac{\alpha}{\beta -1}+ 
         \frac{\alpha^2}{(\beta -1)^3} \right) 
-  \frac{1}{32 \pi^2} \frac{\alpha^2 \beta}{(\beta -1)^3} 
         + {\cal O} (m_\pi^6)  ~,
\eeq
with
\beq
L = \frac{M_R^{d-4}}{16 \pi^2} \left\{ \frac{1}{d-4} 
- \frac{1}{2} \left[ \ln 4\pi - \gamma_E +1 
         \right] \right\}~,
\eeq
and $\gamma_E$ is  the Euler-Mascheroni constant.
We have chosen the regularization scale to be $M_R$.

\medskip\noindent
In fact, the same result for the infrared part can be obtained 
by expanding the baryon propagator in the 
loop integral (a method first used in Ref.~\cite{Ellis:1997kc})
\beq
\int \frac{d^d l}{(2 \pi)^d}  \ \frac{i}{[l^2 - m_\pi^2] \ 
[p^2 + 2 p \cdot l + l^2 - M_N^2]} ~.
\eeq
Counting the loop momentum as $l \sim {\cal O} (m_\pi)$ and taking 
$p^2 \gg (M_N + m_\pi)^2$, one obtains
\beqa
\frac{1}{p^2 - M_N^2} \int \frac{d^d l}{(2 \pi)^d}  \ \frac{i}{l^2 - m_\pi^2}
          \left(  1 - \frac{2 p \cdot l + l^2}{p^2 - M_N^2} +
          \frac{(2 p \cdot l + l^2)^2}{(p^2 - M_N^2)^2} + \ldots \right)~,
\eeqa
which reproduces the result of Eq.~(\ref{eq:IRresult})
after interchanging summation and integration.

\medskip\noindent
After investigating the scalar loop integral one readily obtains the 
infrared singular part of the full one-loop self-energy diagram
with an intermediate pion-nucleon pair (see Fig.~\ref{fig:roperself})
\beqa
\big( \Sigma_{N} (\slashed{p}) \big)_{IR} &=& i \frac{3 g_{NR}^2}{4 f_\pi^2} 
         \int_{IR} \frac{d^d l}{(2 \pi)^d}
   \frac{\slashed{l} (\slashed{p}+ \slashed{l} - M_N) \slashed{l}}{[l^2 - m_\pi^2] 
    \ [(p+l)^2 - M_N^2]}  \no \\
&=&
- \frac{3  g_{NR}^2}{4 f_\pi^2}  \left(  \slashed{p} 
\left[ \frac{m_\pi^4}{32 \pi^2 
  (p^2 - M_N^2)} \ln \left( \frac{m_\pi^2}{M_R^2}\right) + \frac{m_\pi^4}{64 \pi^2 
  (p^2 - M_N^2)} \right] \right. \no \\
&& \qquad \qquad \qquad   \left.
+  \frac{M_N m_\pi^4}{16 \pi^2 
  (p^2 - M_N^2)} \ln \left( \frac{m_\pi^2}{M_R^2}\right) \right)~,
\eeqa
where we have only displayed the finite pieces. The term proportional to $L$ has
been absorbed into the counter terms.
Evaluating the integral at $\slashed{p} = M_R$ yields
\beq \label{eq:resloopN}
\big( \Sigma_N (M_R) \big)_{IR} =
- \frac{3 g_{NR}^2}{256 \pi^2 f_\pi^2 (M_R^2 - M_N^2)}  \ m_\pi^4 \ \left[ 
  (2 M_R+ 4 M_N) \ln \left( \frac{m_\pi^2}{M_R^2}\right) \ 
  + M_R\right]~.
\eeq
Note that this expression preserves both power counting and chiral
symmetry.

\smallskip\noindent
{\bf 3. Pion-Roper loop:}
This corresponds to the standard IR case  and is immediately obtained from
the result in  \cite{Becher:1999he} by replacing $M_N$ by $M_R$ 
(see Fig.~\ref{fig:roperself}),
\beq \label{eq:resloopR}
\big( \Sigma_R (M_R) \big)_{IR} =
- \frac{3 g_R^2}{32 \pi f^2_\pi} m_\pi^3  \left[ 1 + \frac{m_\pi}{2 \pi M_R} 
       + \frac{m_\pi}{2 \pi M_R} \ln \left(  
       \frac{m_\pi^2}{M_R^2}\right) \right] + {\cal O}(m_\pi^5)~.
\eeq
Again, power counting and chiral symmetry are maintained for the IR 
singular part of this diagram.

\smallskip\noindent
{\bf 4. Tadpoles:}
The tadpoles with vertices from $\Lagr_{R}^{(2)} $ yield
\beq  \label{eq:tad}
\Sigma_R^{\rm tad} =
\left( 6 c_1^*  - \frac{3}{4} c_2^* - 3 c_3^* \right) 
    \frac{m_\pi^4}{16 \pi^2 f_\pi^2} \ln \left( \frac{m_\pi^2}{M_R^2}\right)
    + \frac{3}{128 \pi^2 f_\pi^2 } \, c_2^* \, m_\pi^4~.
\eeq
Again, this result agrees with the one for the nucleon by proper substitution
of the LECs and baryon masses.

\smallskip\noindent
{\bf 5. Total Roper self-energy:}
Putting all these pieces together,
 we obtain the following one-loop correction to the Roper mass
\beq
\delta M_R^{\rm (1-loop)} =  \big( \Sigma_N (M_R) \big)_{IR} + \big( \Sigma_R (M_R) \big)_{IR}
                       + \Sigma_R^{\rm tad}  +  \Sigma_R^{\rm tree} 
\eeq
in terms of the renormalized couplings $c_1^*$, $e_1^*$ and the renormalized 
mass $M_R$ for which we use the same notation.

\medskip\noindent
We have not explicitly considered loops with a $\Delta(1232)$-pion
pair\footnote{Note that in the J\"ulich coupled-channels approach, the
Roper is dynamically generated with an important $\pi \Delta$ component
besides the $\sigma N$ one \cite{Krehl:1999km}.}. 
If one treats the $\Delta$
on the same footing as the nucleon field, the contribution will be of the type in 
Eq.~(\ref{eq:resloopN}) and amounts to a renormalization of $g_{NR}$ and $e_1^*$.
However, one must keep in mind that the convergence of the corresponding chiral
series is not as good as for the nucleon due to the smaller
mass difference $M_R^2-M_\Delta^2$.
If, on the other hand, the Delta mass is considered to be of the same size as
the Roper mass, the loop contribution will be similar to the result in 
Eq.~(\ref{eq:resloopR})
and leads to a modification of $g_R$ and the couplings $c_i^*, e_1^*$.
In both scenarios, the effects of the $\Delta\pi$ loop can be absorbed into a redefinition
of the couplings. Since their values are not fixed, we will vary them 
within certain regions, see Sec.~\ref{sec:qmasdep},
such that the inclusion of the $\Delta$ resonance will not alter any of our conclusions.
For a treatment of the $\Delta$ in infrared regularization see, e.g., 
Refs.~\cite{Bernard:2003xf,Bernard:2005fy,Hacker:2005fh}. We also
stress that the explicit inclusion of the $\Delta$ can lead to a complicated
three-small-scales problem (the pion mass and---if considered small---the Roper-Delta  and the
Delta-nucleon splitting) that requires theoretical tools that have not yet been 
developed for baryon chiral perturbation theory.

\section{Quark mass dependence of the Roper mass}  \label{sec:qmasdep}

Before analyzing the pion mass dependence of the Roper mass, we must collect
information on the couplings $g_R$, $g_{NR}$ and the LECs $c_i^*$
$(i=1,2,3)$, and $e_1^*$. 
One obtains $g_{NR} = 0.3 \ldots 0.4$ from a fit to the branching ratio of
the Roper into one pion and a nucleon which is in agreement with the relation
$g_{NR} = \sqrt{R} g_A/2$, where $g_A$ is the axial-vector coupling of the
nucleon and $\sqrt{R} = 0.53 \pm 0.04$ (for details, see Ref.~\cite{Bernard:1995gx}).
For $g_R$ the naive quark model predicts $g_R = g_A$, and we set here $g_R = 1.0$
so that $g_A$ and $g_R$ are roughly of the same size, see also \cite{Hernandez:2002xk}. 

\medskip\noindent
To leading order in the chiral expansion,
the LEC $c_1^*$ measures the $\sigma$-term in the Roper state and it is thus
bounded from above by the value of the pion-nucleon $\sigma$-term. This
means $|c_1^*| \lesssim 1\,$GeV$^{-1}$. More realistically, a natural
value for $c_1^*$ would be around $-0.5\,$GeV$^{-1}$ because  $\sigma$-terms
are expected to become smaller with the resonance excitation energy (see
also the related discussion on the $\pi \Delta$  $\sigma$-term in
Refs.~\cite{Bernard:2005fy,Holl:2005st,Cavalcante:2005mb,Hacker:2005fh}). The
sign of $c_1^*$ should be negative since the quark masses contribute
positively to the hadron masses. The nucleon LECs $c_2$ and $c_3$ are much 
enhanced compared to the natural values
$|c_i| \lesssim 1\,$GeV$^{-1}$ because of the nearby and strongly coupled delta
resonance \cite{Bernard:1996gq}. This is not expected to be the case for the Roper
resonance. Consequently, the LECs $c_{2,3}^*$ can be bounded conservatively 
by $\pm 1\,$GeV$^{-1}$. The pion decay constant in the chiral limit
is taken to be $f_\pi = 87$ MeV \cite{Colangelo:2003hf}.

\begin{figure}[t]
\centering
\includegraphics[width=0.7\textwidth]{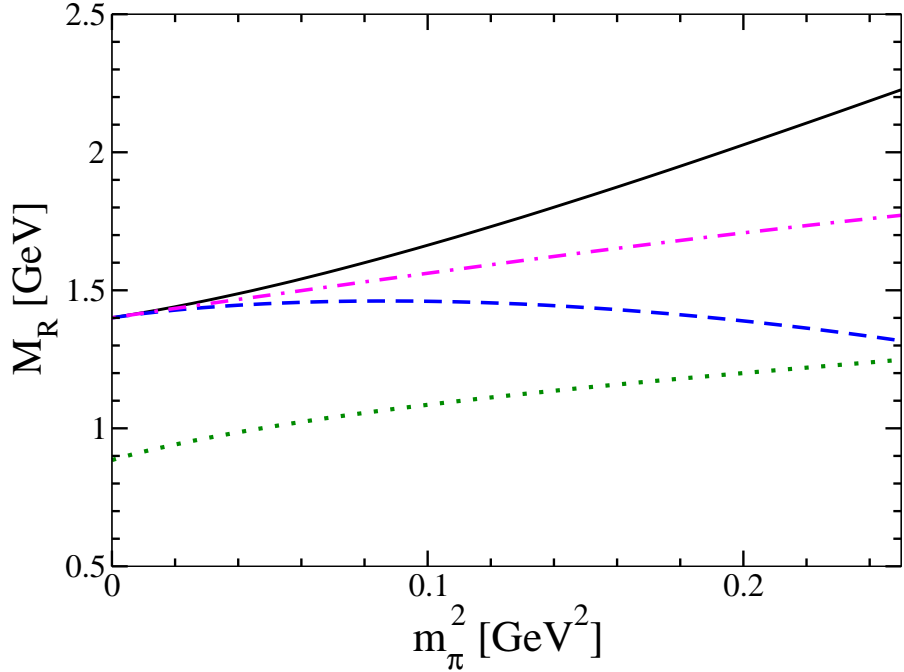} 
\caption{Quark mass dependence of the Roper mass for different parameter sets 
$c_1^* = -0.5, c_{2,3}^*, e_1^*$. The $c_i$ are in units of GeV$^{-1}$ and 
$e_1$ is given in GeV$^{-3}$.  
and couplings $g_R = 1.0, g_{NR} = 0.35$. The solid curve corresponds to
$c_2^* = 1.0, c_3^* = 1.0, e_1^* = 0.5$, the dashed one to 
$c_2^* = -1.0, c_3^* = -1.0, e_1^* = -0.5$ and the dot--dashed one to
$c_2^* = c_3^* = e_1^* = 0$. The dotted curve represents the quark mass 
dependence of the nucleon, see Ref.~\cite{Bernard:2003rp}. The values of the
corresponding LECs are: $c_1 = -0.9, c_2 = 3.2, c_3 = -3.45,e_1 = -1.0$.
}
\label{fig:mro}
\end{figure}

\medskip\noindent
In Fig.~\ref{fig:mro} an estimated range for the pion mass dependence of the Roper mass
is presented by taking the extreme values for $c_{2,3}^*$ and $e_1^*$, while keeping
$c_1^* = -0.5\, {\rm GeV}^{-1}, g_{NR} = 0.35, g_R = 1$ fixed. 
The masses of the baryons in the chiral limit
are taken to be $M_N = 0.885\,$GeV \cite{Bernard:2003rp}
and $M_R = 1.4\,$GeV, respectively.
The dash-dotted curve is obtained by setting the couplings $c_{2,3}^*,e_1^*$
all to zero, and exhibits up to an offset
a similar quark mass dependence as the nucleon result (dotted curve,
taken from Ref.~\cite{Bernard:2003rp}). 
It should be emphasized, however, that the one-loop formula cannot be trusted
for pion masses much beyond 350 MeV.
For similar results for the
nucleon mass, see also Ref.~\cite{Procura:2006bj}.

\medskip\noindent
In the numerical calculation we have employed the pion decay constant 
in the chiral limit,
$f_\pi = 87$ MeV. However, we could have equally well used the physical pion
decay constant, $F_\pi = 92.4$ MeV, as the difference in the chiral expansions
appears either at chiral order ${\cal O}(m_\pi^6)$ in 
Eqs.~(\ref{eq:resloopN}, \ref{eq:tad})
or at order ${\cal O}(m_\pi^5)$ for the Roper-pion loop which is beyond
the accuracy of the present investigation. The numerical results for these
contributions would change by the amount of $f_\pi^2/F_\pi^2 \approx 0.89$
and do not lead to significant changes in the results. Stated differently, the replacement of
$f_\pi$ by $F_\pi$ in these formulae induces a correction due to the
quark mass expansion of $F_\pi$
\beq
F_\pi = f_\pi \left( 1 + \frac{m_\pi^2}{16 \pi^2 f_\pi^2} \bar{l}_4 + {\cal O}(m_\pi^4)   \right)
\eeq
with $\bar{l}_4 = 4.33$. The modifications at leading order are then
\beq 
- \frac{3 g_{NR}^2}{2048 \pi^4 F_\pi^4 (M_R^2 - M_N^2)} \ \bar{l}_4 \ m_\pi^6 \ \left[ 
  (2 M_R+ 4 M_N) \ln \left( \frac{m_\pi^2}{M_R^2}\right) \ 
  + M_R\right]~
\eeq
for the nucleon-pion loop,
\beq \
- \frac{3 g_R^2}{256 \pi^3 F^4_\pi} \ \bar{l}_4 \ m_\pi^5  
\eeq
for the Roper-pion loop, and
\beq  
\left( 6 c_1^*  - \frac{3}{4} c_2^* - 3 c_3^* \right) \ \bar{l}_4 \
    \frac{m_\pi^6}{128 \pi^4 F_\pi^4} \ln \left( \frac{m_\pi^2}{M_R^2}\right)
    + \frac{3}{1024 \pi^4 F_\pi^4 } c_2^* \ \bar{l}_4 \ m_\pi^6~.
\eeq
for the tadpoles. These corrections at higher chiral orders are indeed small
and can be safely neglected for small pion masses. In fact, the variations
induced by these corrections are within the band for $M_R (m_\pi^2)$
given in Fig.~\ref{fig:mro}. 

\medskip\noindent
In this work, we have calculated the chiral corrections to the
Roper mass  to one-loop order.
The approach is based on an extension of infrared regularization
which allows for the unambiguous isolation of the infrared singular part of the
loops stemming from the pion poles.
At the same time, chiral symmetry is preserved and 
a chiral counting scheme emerges. 
The considered Feynman diagrams contain two different heavy mass scales
$M_N, M_R$ which we consider to satisfy the relation $M_N^2 \ll M_R^2$.
The utilized formalism is in general suited to study systems with two heavy
mass scales in addition to a light mass scale. In this sense,
it can be applied
to other resonances as well, such as the $S_{11}(1535)$. In this case, however,
an SU(3) calculation is necessary due to the important $\eta N$ decay channel.

\bigskip

\section*{Acknowledgements}
We thank J\"urg Gasser for an informative discussion on sigma terms and
Siegfried Krewald for comments on the structure of the Roper.

\newpage


\end{document}